\documentclass[conference]{IEEEtran}
\usepackage[dvips]{graphicx}
\usepackage{cite,amssymb,amsmath,epsfig,color}
\newcommand{\beq}{\begin{equation}}
\newcommand{\eeq}{\end{equation}}
\newcommand{\beqn}{\begin{eqnarray}}
\newcommand{\eeqn}{\end{eqnarray}}
\DeclareMathOperator*{\argmin}{arg\,min}

\long\def\symbolfootnote[#1]#2{\begingroup%
\def\thefootnote{\fnsymbol{footnote}}\footnote[#1]{#2}\endgroup}

\title{Adaptive ADMM in Distributed Radio Interferometric Calibration}
\author{\IEEEauthorblockN{Sarod Yatawatta\IEEEauthorrefmark{1}}
\IEEEauthorblockA{ASTRON,\\ The Netherlands Institute for Radio Astronomy,\\Dwingeloo, The Netherlands\\
Email: yatawatta@astron.nl\IEEEauthorrefmark{1}}
\and
\IEEEauthorblockN{Faruk Diblen and Hanno Spreeuw}
\IEEEauthorblockA{Netherlands eScience Center\\ Science Park 140 (Matrix I)\\ 1098 XG Amsterdam, The Netherlands\\
}
}

\begin{document}
%\ninept
%
\maketitle
\begin{abstract}
Distributed radio interferometric calibration based on consensus optimization has been shown to improve  the estimation of systematic errors in radio astronomical observations. The intrinsic continuity of systematic errors across frequency is used by a consensus polynomial to penalize traditional calibration. Consensus is achieved via the use of alternating direction method of multipliers (ADMM) algorithm. In this paper, we extend the existing distributed calibration algorithms to use ADMM with an adaptive penalty parameter update. Compared to a fixed penalty, its adaptive update has been shown  to perform better in diverse applications of ADMM. In this paper, we compare two such popular penalty parameter update schemes: residual balance penalty update and spectral penalty update (Barzilai-Borwein). We apply both schemes to distributed radio interferometric calibration and compare their performance against ADMM with a fixed penalty parameter. Simulations show that both methods of adaptive penalty update improve the convergence of ADMM but the spectral penalty parameter update shows more stability.
\end{abstract}
\section{Introduction}
Raw data produced by radio interferometric arrays are almost always corrupted by systematic errors introduced by the propagation medium (atmosphere) and by the instrument (beam and receiver). Calibration is estimation of such errors and correcting the data to remove the effects of such errors. In order to handle large volumes of data produced by modern radio interferometric arrays, efficient and accurate calibration algorithms are necessary. The recent surge in popularity of distributed optimization algorithms \cite{BT,boyd2011} have enabled us to address some of these issues related to calibration of large data volumes in radio astronomy. 
 
With the use of consensus optimization \cite{boyd2011}, it has been shown that \cite{DCAL,Brossard2016} distributed calibration provides a way of distributing the computational burden over a network of computers while at the same time improving the quality of calibration. This enables processing of huge amounts of data that are already stored at various locations across a network of computers and using the local computational power available at each particular location with minimal network communication. In order to do this, the inherent continuity of systematic errors over frequency is exploited and this is added as a constraint onto calibration. With this modification, calibration  is transformed into a consensus optimization \cite{boyd2011} problem and we use alternating direction method of multipliers (ADMM) \cite{BT}  as the underlying algorithm to reach consensus. Similar distributed computing strategies are being developed in radio interferometric imaging as well \cite{Meil2016,Degu2016,Onose2016,Onose2017}.

There is widespread use of the ADMM algorithm in various and diverse applications including machine learning \cite{boyd2011,yang2017admm}, image processing \cite{setzer2011operator} and medical imaging \cite{Gong17}. Unlike most applications, radio interferometric calibration using ADMM has some unique properties. The cost function that is minimized in calibration is nonlinear and nonconvex. Even though the systematic errors (e.g., ionosphere, beam shape, receiver gain) have well behaved continuity, the exact description of this behavior using polynomials of low degree is not accurate enough. This is due to the complex interactions of the systematic errors, especially when a wide area of the sky is observed. Therefore, consensus can be achieved only based on an approximate model, which is clearly different than most other application. Furthermore, while most other applications use complicated network topologies, we use a simpler topology with  a set of data processing nodes connected to one fusion center.

The convergence rate of ADMM and its dependence on the penalty parameter are well studied and generally with proper initialization \cite{Giselsson,nishihara2015general,Teix2016,Ghadimi2015,Hong15} and also with adaptive update \cite{boyd2011,He2000,Wohlberg2017,Zheng2016,Zheng2016NIPS,Zheng2017}, it has been shown that convergence could be improved. Our previous work \cite{EUSIPCO2016} focused on the initialization of the penalty parameter using the smallest eigenvalue of the Hessian of the cost function. In this paper, we further improve our calibration by enabling adaptive update of the penalty parameter. We compare two adaptive penalty parameter update schemes in this paper. The first scheme is {\em residual balancing} \cite{He2000,boyd2011,Wohlberg2017} where the penalty is updated to balance the primal and dual residuals of the ADMM algorithm. In contrast, the second method is based on the {\em spectral} penalty parameter update \cite{Zheng2016,Zheng2016NIPS,Zheng2017}, which is based on the Barzilai-Borwein adaptive step size selection method \cite{BB1988} in gradient descent. We compare both methods using simulated radio interferometric data and both methods give better performance than the case where the penalty parameter is kept fixed. Radio astronomical observations (even made by the same interferometric array) have varying instrumental models depending on the direction of the sky being observed and the frequency range and time interval at which data is taken. Therefore, we find that the residual balancing method needs proper tuning \cite{Wohlberg2017} to suit each particular observation. On the other hand, the spectral penalty update is less prone to changes in instrumental model and is better suited in processing large data volumes with minimal manual intervention.   

The rest of the paper is organized as follows: In section \ref{sec:calib} we give an overview of distributed radio interferometric calibration using consensus optimization. In section \ref{sec:update}, we present schemes for updating  the penalty parameter with ADMM iterations. Simulation results are presented in section \ref{sec:results} where we demonstrate the improved performance with an adaptive penalty parameter. Finally, we draw our conclusions in section \ref{sec:conclusions}.

Notation: Matrices and vectors are denoted by bold upper and lower case letters as ${\bf J}$ and ${\bf v}$, respectively. The transpose and the  Hermitian transpose are given by $(.)^T$ and $(.)^H$. The matrix  Frobenius norm is given by $\|.\|$. The set of real and complex numbers are denoted by  ${\mathbb R}$ and ${\mathbb C}$ respectively. The identity matrix is given by $\bf I$. The matrix trace operator is given by $\rm{trace}(.)$. 
\section{Radio Interferometric Calibration}\label{sec:calib}
We consider an interferometric array consisting of $N$ stations. The radiation originating from any given direction in the sky is seen at the interferometer formed by stations $p$ and $q$ as ${\bf V}_{pqf}$ \cite{HBS}
\beq \label{ME}
{\bf V}_{pqf}={\bf J}_{pf} {\bf C}_{pqf} {\bf J}_{qf}^H + {\bf N}_{pqf}
\eeq
where $f$  is the frequency at which data is taken and  ${\bf V}_{pqf},{\bf J}_{pf},{\bf J}_{qf},{\bf C}_{pqf},{\bf N}_{pqf}\in \mathbb{C}^{2\times 2}$. The error-free signal from the sky is given by ${\bf C}_{pqf}$, which is essentially the Fourier transform of the sky model and can be pre-computed \cite{TMS}. The systematic errors corrupting the true signal are given by the Jones matrices ${\bf J}_{pf},{\bf J}_{qf}$. The noise is given by ${\bf N}_{pqf}$ and is assumed to have complex, zero mean, circular Gaussian elements.

In reality, the observed data is the sum of many signals as in (\ref{ME}), originating from many directions in the sky. However, we can simplify calibration along many such directions in the sky using the space alternating generalized expectation maximization (SAGE) algorithm \cite{Fess94,Kaz2}. In this paper, we describe our algorithms for calibration along a single direction but the results presented in section \ref{sec:results} is based on calibration along multiple directions.

The cost function that is minimized is given as
\beq \label{cost}
g_{f}({\bf J}_f)= \sum_{p,q}\| {\bf V}_{pqf} - {\bf A}_p{\bf J}_f {\bf C}_{pqf} ({\bf A}_q{\bf J}_f)^H \|^2
\eeq
where the systematic errors for all $N$ stations are grouped as ${\bf J}_f\in \mathbb{C}^{2N\times 2}$,
\beq
{\bf J}_f\buildrel\triangle\over=[{\bf J}_{1f}^T,{\bf J}_{2f}^T,\ldots,{\bf J}_{Nf}^T]^T.
\eeq
Using the canonical selection matrix ${\bf A}_p$ ($\in \mathbb{R}^{2\times 2N}$), where only the $p$-th block is ${\bf I} \in \mathbb{R}^{2\times 2}$, 
\beq \label{Ap}
{\bf A}_p \buildrel\triangle\over=[{\bf 0},{\bf 0},\ldots,{\bf I},\ldots,{\bf 0}],
\eeq
we can select the systematic errors for the station $p$  as ${\bf A}_p{\bf J}_f$.   
Note that in (\ref{cost}), the summation is taken over all the baselines $pq$ that have data, within a small bandwidth and time interval within which the systematic errors are assumed to be fixed.

The variation of ${\bf C}_{pqf}$ in (\ref{cost}) with $f$ is smooth and is known. Moreover, the variation of the systematic errors ${\bf J}_f$ is assumed smooth, but not exactly known. While conventional calibration minimizes (\ref{cost}) without exploiting this smoothness, we can add this as an additional constraint. Given that data is collected at a set of frequencies $\mathcal{F}=\{f_1,f_2,\ldots,f_P\}$, the reformulated calibration problem can be stated as \cite{DCAL}
\beq \label{prob}
{\bf J}_f = \underset{{\bf J}}{\argmin}\ g_{f}({\bf J})\ \mathrm{subject\ to}\ {\bf J}_f={\bf B}_f {\bf Z},\  \forall f\in\mathcal{F}.
\eeq
In (\ref{prob}), the constraint ${\bf J}_f={\bf B}_f {\bf Z}$ enforces the continuity of ${\bf J}_f$ with $f$. In order to do this, we use ${\bf B}_f \in \mathbb{R}^{2N\times 2NF}$ which is a set of $F$ basis functions in frequency (we use the same basis functions for all $N$ stations) and ${\bf Z} \in \mathbb{C}^{2NF\times 2}$ is the global variable that enforces continuity across all frequencies $f\in\mathcal{F}$.

We can transform (\ref{prob}) into a consensus optimization problem as follows. First, we create the augmented Lagrangian as
\beq \label{aug}
L_f({\bf J}_f,{\bf Z},{\bf Y}_f)=g_{f}({\bf J}_f) + \|{\bf Y}_f^H({\bf J}_f-{\bf B}_f {\bf Z})\| + \frac{\rho_f}{2} \|{\bf J}_f-{\bf B}_f {\bf Z}\|^2
\eeq
where the subscript $(.)_f$ denotes data (and parameters) at frequency $f$. In (\ref{aug}), $g_{f}({\bf J}_f)$ is the original cost function as in (\ref{cost}). The Lagrange multiplier is given by ${\bf Y}_f$ ($\in \mathbb{C}^{2N\times 2}$). The global variable ${\bf Z}$ is shared by data at all $P$ frequencies. One noteworthy difference from our previous work \cite{DCAL,EUSIPCO2016} is that rather than being fixed, the penalty parameter $\rho_f$ is variable.

The ADMM iterations $n=1,2,\ldots$ for solving (\ref{aug}) are given as
\beqn \label{step1}
({\bf J}_f)^{n+1}= \underset{{\bf J}}{\argmin}\ \ L_f({\bf J},({\bf Z})^n,({\bf Y}_f)^n,\rho_f^n)\\ \label{step2}
({\bf Z})^{n+1}= \underset{{\bf Z}}{\argmin}\ \ \sum_f L_f(({\bf J}_f)^{n+1},{\bf Z},({\bf Y}_f)^n,\rho_f^n)\\ \label{step3}
({\bf Y}_f)^{n+1}=({\bf Y}_f)^n + \rho_f^n\left( ({\bf J}_f)^{n+1}-{\bf B}_f ({\bf Z})^{n+1} \right)\\ \label{step4}
\rho_f^{n+1}=\mathrm{update\ penalty\ parameter} 
\eeqn
where we use the superscript $(.)^n$ to denote the $n$-th iteration where (\ref{step1}) to (\ref{step4}) are executed in order. The steps (\ref{step1}),(\ref{step3}) and (\ref{step4}) are done for each $f$ in parallel, at each compute (slave) node. The slave nodes are distributed across a network of computers. The update of the global variable in (\ref{step2}) is done at the fusion center. The extra step (\ref{step4}), which is an improvement from our previous work \cite{DCAL,EUSIPCO2016}, will be discussed in section \ref{sec:update}.

\section{Updating penalty parameter}\label{sec:update}
 Since the constraint ${\bf J}_f={\bf B}_f {\bf Z}$ in (\ref{prob}) is not guaranteed to be entirely accurate, increasing the value of $\rho_f$ too much would bias the solutions towards this constraint and therefore increase the estimation error. Moreover, the observed data ${\bf V}_{pqf}$ in (\ref{ME}) always contain contributions from weak signals from the sky that are not part of the sky model ${\bf C}_{pqf}$ \cite{Kaz3,SIRP} and we do not expect to see continued improvement of solutions with increased number of ADMM iterations. Hence, in this paper, we use a fixed number of ADMM iterations rather than using various stopping criteria \cite{boyd2011}.

We compare two popular schemes for the update of the penalty parameter, which can be plugged in to (\ref{step4}) of the ADMM iterations. In both cases, the initial value ($n=1$) for $\rho_f^1$ is chosen by using the magnitude of the lowest eigenvalue of the Hessian, say $|\lambda|$, (scaled down by a factor $\approx 1/10$) as described in \cite{EUSIPCO2016}. In addition, to safeguard that the updated $\rho_f^n$ does not increase too much, thereby giving too much weight to the constraint in (\ref{aug}), $|\lambda|$ is also used as an upper bound to all updates of $\rho_f$. In other words, if a possible update of $\rho_f$ is higher than $|\lambda|$, it is clamped at this value.

\subsection{Residual balancing penalty update}
The idea behind the residual balancing method \cite{boyd2011,He2000,Wohlberg2017} is to select penalty parameter such that both the primal residual ${\bf R}_f^n$
\beq \label{primal}
{\bf R}_f^n={\bf J}_f^n - {\bf B}_f {\bf Z}^n
\eeq
and the dual residual ${\bf S}_f^n$
\beq \label{dual}
{\bf S}_f^n=\rho_f^{n}{\bf B}_f ({\bf Z}^n -{\bf Z}^{n-1})
\eeq
have balanced norms. This provides a balance between the original cost function and the constraint in (\ref{aug}). Heuristically, the penalty parameter is updated as
\beq
\rho_f^{n+1}=\begin{cases}
\tau \rho_f^n & \mathrm{if}\ \|{\bf R}_f^n\| > \mu \|{\bf S}_f^n\|\\
\tau^{-1} \rho_f^n & \mathrm{if}\ \|{\bf R}_f^n\| < \mu^{-1} \|{\bf S}_f^n\|\\
\rho_f^n & \mathrm{otherwise,}
\end{cases}
\eeq
where $\mu (>1)$ and $\tau (>1)$ are two constants that are given a priori and typical values used are $\mu=10$ and $\tau=2$ \cite{boyd2011}.
\subsection{Spectral penalty update}
The spectral parameter update \cite{Zheng2016,Zheng2016NIPS} is based on the Barzilai-Borwein method \cite{BB1988} used in adaptive step size selection of gradient descent optimization schemes \cite{Zhou2006}. 
For this scheme, we need extra variables that have the lifetime of the total ADMM iterations, i.e., $ \widehat{\bf Y}_f^0,\widehat{\bf Y}_f,{\bf J}_f^0 \in \mathbb{C}^{2N\times 2}$. At the first ADMM iteration ($n=1$), using the current solutions $\left({\bf J}_f\right)^1$, initialize $\widehat{\bf Y}_f^0={\bf J}_f^0=\left({\bf J}_f\right)^1$. It is also noteworthy that the penalty is not updated at each ADMM iteration, on the contrary, it is done with a periodicity $T (\ge2)$.
At the $n$-th ADMM iteration, if $n$ is a multiple of $T$, we perform an update as follows. First, we find step sizes $\alpha_{SD},\alpha_{MG}$ (the subscripts $SD$ and $MG$ stand for steepest descent and minimum gradient \cite{Zhou2006}) and correlation coefficient $\alpha$ as
\beq \label{stepA1}
\left(\widehat{\bf Y}_f\right)^{n+1}=({\bf Y}_f)^n + \rho_f^n\left( ({\bf J}_f)^{n+1}-{\bf B}_f ({\bf Z})^{n} \right),
\eeq
\beq  \label{stepA2}
\Delta {\bf Y}_f = \left(\widehat{\bf Y}_f\right)^{n+1} - \widehat{\bf Y}_f^0,\ 
 \Delta {\bf J}_f =\left({\bf J}_f\right)^{n+1}-{\bf J}_f^0,
\eeq
\beqn \label{stepA3}
\delta_{11}&=&\mathrm{trace}\left(\mathrm{real}(\Delta{\bf Y}_f^H \Delta{\bf Y}_f)\right)\\\nonumber
\delta_{12}&=&\mathrm{trace}\left(\mathrm{real}(\Delta{\bf Y}_f^H \Delta{\bf J}_f)\right)\\\nonumber 
\delta_{22}&=&\mathrm{trace}\left(\mathrm{real}(\Delta{\bf J}_f^H \Delta{\bf J}_f)\right),
\eeqn
\beq \label{stepA4}
\alpha=\frac{\delta_{12}}{\sqrt{\delta_{11} \delta_{22}}},\ \alpha_{SD}=\frac{\delta_{11}}{\delta_{12}},\ \alpha_{MG}=\frac{\delta_{12}}{\delta_{22}}.
\eeq
Note that (\ref{stepA1}) differs from (\ref{step3}) because $({\bf Z})^{n}$ is used in the former and $({\bf Z})^{n+1}$ is used in the latter.
Next, a candidate for the updated penalty $\hat{\alpha}$ is chosen as
\beq\label{stepA5}
\hat{\alpha}=\begin{cases}
\alpha_{MG} & \mathrm{if}\  2\alpha_{MG}> \alpha_{SD}\\
\alpha_{SD}-\frac{\alpha_{MG}}{2}\ \mathrm{otherwise}.
\end{cases}
\eeq
Finally, if there is sufficient correlation for this update, 
\beq \label{stepA6}
\rho_f^{n+1}=\begin{cases}
\hat{\alpha} & \mathrm{if}\ \alpha\ge \underline{\alpha}\\
\rho_f^n & \mathrm{otherwise}
\end{cases}
\eeq
 where $\underline{\alpha}$ is a constant (typically $>0.2$) to ensure the new update is not a spurious result. At the last step, the auxiliary variables are updated as
\beq \label{stepA7}
 \widehat{\bf Y}_f^0= \left(\widehat{\bf Y}_f\right)^{n+1},\ {\bf J}_f^0=\left({\bf J}_f\right)^{n+1}
\eeq
to be used in the next update of the penalty.

We compare the performance of both aforementioned methods in section \ref{sec:results}. Comparing the computational cost, the residual balancing method is simpler but the spectral method is also not expensive as all what is needed is a few linear operations and inner products. Moreover, the spectral update is not performed at every ADMM iteration.

\section{Simulation Results}\label{sec:results}
We simulate an array of $N=47$ receivers that calibrate along $K=4$ directions in the sky. The settings for the simulations are quite similar to the one used in \cite{EUSIPCO2016}. The matrices corresponding to the systematic errors, i.e., ${\bf J}_{pk},{\bf J}_{qk}$ in (\ref{ME}) are generated with their elements having values drawn from a complex uniform distribution in $[0,1]$, multiplied by a frequency dependence given by a random $8$-th order ordinary polynomial in frequency. The intensities of the $K=4$ sources are randomly generated in the range $[1,5]$ intensity units and their positions are randomly chosen in a field of view of about $7\times 7$ square degrees. The variation of intensities with frequency is given by a power law with randomly generated exponent in $[-1,1]$. The noise matrices ${\bf N}_{pq}$ in (\ref{ME}) are simulated to have complex circular Gaussian random variables. The variance of the noise is changed according to the signal to noise ratio ($\rm{SNR}=30$) 
\beq
\mathrm{SNR}\buildrel\triangle\over=\frac{\sum_{p,q} \|{\bf V}_{pq}\|^2}{\sum_{p,q} \| {\bf N}_{pq}\|^2}.
\eeq
In addition, we add the signals of $400$ weak sources, with intensities uniformly distributed in $[0.01,0.1]$ intensity units, randomly located within the $7\times 7$ square degrees field of view. The signals of these $400$ weak sources act as an additional source of noise and are simulated without any systematic errors.

We generate data for $P=8$ frequency channels in the range $115$ to $185$ MHz. For calibration, we setup a $3$-rd order polynomial model ($F=4$), using Bernstein basis functions \cite{Farouki} for the matrix ${\bf B}_f$ in (\ref{aug}). We intentionally use a lower order frequency dependence than what is actually present in the data to create a realistic scenario where the exact model for the systematic errors is not known. Initial values for the calibration parameters are always set as ${\bf J}_p={\bf I}$ for $p\in[1,N]$. The penalty parameter for the $K$ directions are initialized using $1/10$ of the magnitude of the lowest eigenvalue of the Hessian of (\ref{cost}), as described in \cite{EUSIPCO2016}. The magnitude of the lowest eigenvalue is also used as an upper bound to any updated value of the penalty. 

The accuracy of calibration is measured using the normalized (and averaged over all directions) mean squared error (NMSE) between true ${\bf J}_f$ and its estimate as
\beq\label{nmse}
\mathrm{NMSE}\buildrel\triangle\over=\frac{1}{\sqrt{2KN}}\sqrt{ \sum_{\mathrm{over\ all\ } k} \|{\bf {J}}_f-\widehat{\bf {J}}_f {\bf {U}}\|^2}
\eeq
where ${\bf U}$ is a unitary matrix that removes the unitary ambiguity in the estimated $\widehat{\bf J}_f$ \cite{interpolation}.

\begin{figure}[htbp]
\begin{minipage}[b]{0.98\linewidth}
\centering
\centerline{\epsfig{figure=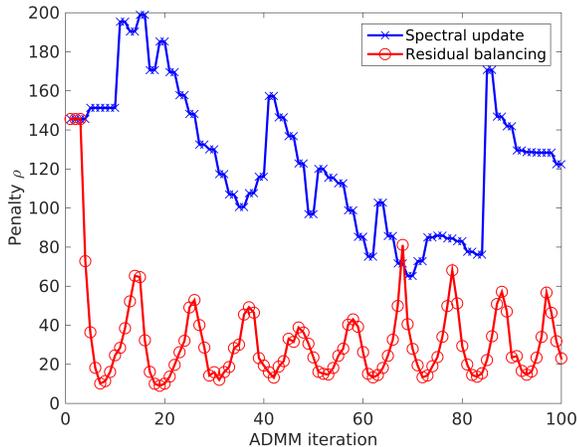,width=8.5cm}}
\end{minipage}
\caption{Variation of  $\rho_f$ for the two adaptive update schemes with ADMM iteration.}
\label{figrho}
\end{figure}

We compare the spectral penalty update (with $T=2$ and $\underline{\alpha}=0.2$), and the residual balance penalty update (with $\mu=20$ and $\tau=2$) against the performance of ADMM with a fixed penalty. In Fig. \ref{figrho}, we show the variation of $\rho_f$ with ADMM iterations for both adaptive schemes, for one direction (out of $K$) and frequency.

\begin{figure}[htbp]
\begin{minipage}[b]{0.98\linewidth}
\centering
\centerline{\epsfig{figure=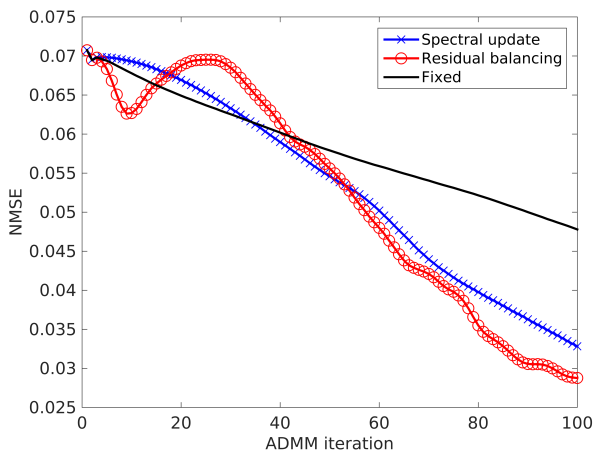,width=8.5cm}}
\end{minipage}
\caption{NMSE (averaged over all frequencies) variation, with ADMM iteration.} \label{fignmseall}
\end{figure}

In Fig. \ref{fignmseall}, we show the variation of NMSE (averaged over all frequencies) with ADMM iterations, for both adaptive penalty updated schemes as well as for the case where the penalty is fixed. In all cases, the initial penalty parameter is the same. We see that while both adaptive update schemes perform better than ADMM with a fixed penalty, the residual balance update scheme shows more oscillations.  

\begin{figure}[htbp]
\begin{minipage}[b]{0.98\linewidth}
\centering
\centerline{\epsfig{figure=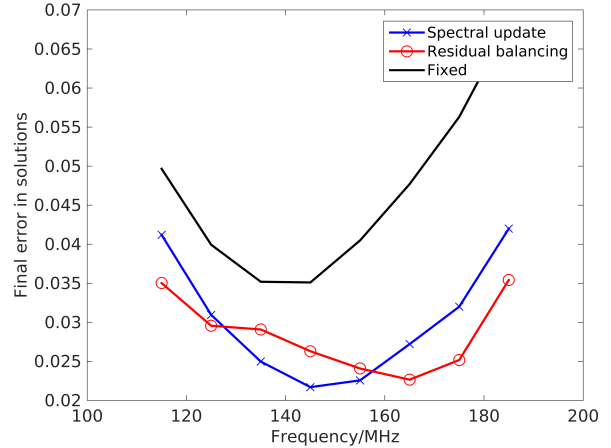,width=8.5cm}}
\end{minipage}
\caption{Final NMSE after $100$ ADMM iterations.} \label{fignmsefinal}
\end{figure}

In Fig. \ref{fignmsefinal}, we show the final NMSE (after $100$ ADMM iterations) for all $P=8$ frequencies. Both adaptive penalty update methods have better NMSE than ADMM with a fixed penalty. In a real observation, data is calibrated by taking small segments and for a full observation, calibration such as the one simulated in this example has to be carried out thousands of times. Hence, rather than continuing ADMM iterations until convergence, the iterations are halted at a predefined value. Therefore, calibration schemes that show less oscillations in NMSE are better suited.  Therefore, we give preference to the spectral penalty update as a practical calibration scheme.

\section{Conclusions}\label{sec:conclusions}
We have compared the performance of adaptive penalty update in ADMM applied to distributed radio interferometric calibration. We used two popular penalty update schemes, namely the residual balancing update  and the spectral (Barzilai-Borwein) update. Both methods improve performance of calibration compared with ADMM with a fixed penalty parameter. However, the spectral penalty update shows more stability and is preferable in practical applications.  Software for distributed radio interferometric calibration with the spectral penalty parameter update is available at http://sagecal.sf.net/ and https://github.com/nlesc-dirac/sagecal.

\section*{Acknowledgment}
This work is supported by Netherlands eScience Center (project DIRAC, grant 27016G05) and the European Research Council (project LOFARCORE, grant 339743).

\bibliographystyle{IEEE}
\bibliography{references}

\end{document}